\documentclass[conference]{IEEEtran}
\IEEEoverridecommandlockouts
\usepackage{cite}
\usepackage{booktabs}
\usepackage{amsmath,amssymb,amsfonts}
\usepackage{algorithmic}
\usepackage{graphicx}
\usepackage{textcomp}
\usepackage{xcolor}
\def\BibTeX{{\rm B\kern-.05em{\sc i\kern-.025em b}\kern-.08em
    T\kern-.1667em\lower.7ex\hbox{E}\kern-.125emX}}

\begin{document}

\title{A deep learning based multiscale approach to segment cancer area in liver whole slide image\\}

\author{\IEEEauthorblockN{Yanbo Feng}
\IEEEauthorblockA{\textit{INSA Centre Val de Loire} \\
\textit{Universit\'e d’Orl\'eans, PRISME}\\
Bourges, France \\
yanbo.feng@insa-cvl.fr}
\and
\IEEEauthorblockN{Adel Hafiane}
\IEEEauthorblockA{\textit{INSA Centre Val de Loire} \\
\textit{Universit\'e d’Orl\'eans, PRISME}\\
Bourges, France \\
adel.hafiane@insa-cvl.fr}
\and
\IEEEauthorblockN{Hélène Laurent}
\IEEEauthorblockA{\textit{INSA Centre Val de Loire} \\
\textit{Universit\'e d’Orl\'eans, PRISME}\\
Bourges, France \\
helene.laurent@insa-cvl.fr}
}

\maketitle

\begin{abstract}
This paper addresses the problem of liver cancer segmentation in Whole Slide Image (WSI).
We propose a multi-scale image processing method based on automatic end-to-end deep neural network algorithm for segmentation of cancer area.
A seven-levels gaussian pyramid representation of the histopathological image was built to provide the texture information in different scales.
In this work, several neural architectures were compared using the original image level for the training procedure. 
The proposed method is based on U-Net applied to seven levels of various resolutions (pyramidal subsumpling). 
The predictions in different levels are combined through a voting mechanism. 
The final segmentation result is generated at the original image level. 
Partial color normalization and weighted overlapping method were applied in preprocessing and prediction separately.
The results show the effectiveness of the proposed multi-scales approach achieving better scores compared to the state-of-the-art. 
\end{abstract}


\section{Introduction}
The liver is a visceral organ most often involved in the metastatic spread of cancer. For the best practice, early diagnosis of liver cancer is important but many people don't even know that they have hepatitis. Hepatocellular Carcinoma (HCC) represents about 90\% of primary liver cancers and constitutes a major global health problem\cite{challenge}. The incidence of HCC is increasing worldwide; it is amongst the leading causes of cancer mortality globally. Between 1990 and 2015 newly diagnosed HCC cases increased by 75\%, mainly due to changing age structures and population growth \cite{challenge}.
Histopathology image analysis is a medical discipline which completely provides direct and reliable information to diagnose diseases. 
Since the early 21th century, the evolution of electronic equipment, which allows entire slides to be imaged and permanently stored at high resolution, has largely contributed to the prosperous development of whole slide imaging (WSI) for computer-assisted diagnosis (CAD). 

With the development of computer vision and artificial intelligence techniques, CAD is widely applied in diagnosis to assist medical experts to interpret medical images.
Digital pathology (DP) is the process by which histology slides are digitized to produce high resolution images via whole slide digital scanners \cite{Gurcan2009Histopathological}.
As an important part of CAD, its aim is to acquire, manage and interpret pathology information generated from digitized glass slides \cite{Chen2017DCAN}.
Although the quantity and complexity of information present in the images, several applications of DP have been studied broadly such as segmentation of desired regions or objects\cite{Hong2010Segmentation}, counting normal or cancer cells\cite{2019U}, recognizing tissue structures\cite{Veronika2019Algorithm}, grading and prognosis of cancers\cite{Wei2019Pathologist}, etc.
It has capacity to effectively facilitate the processing and the analysis of the medical information in the sense of the expert, decrease the intense workload of highly specialized pathologists, and increase the level of inter-observer agreement\cite{Tang2009Computer}. 

WSI is an important source of information, it contains region and cellular-level features, which have significant implications for diagnosis. 
Its complexity is higher than many imaging modalities because of its large size (a resolution of 100k$\times$100k is common) and presence of color information (haematoxylin and eosin, immunohistochemistry). 
The tissue collected during the biopsy is commonly stained with hematoxylin and eosin (H\&E) prior to the visual analysis by the specialists. 
The staining enhances nuclei (purple) and cytoplasm (pinkish), as well as other structures of interest \cite{Brogi2001Rosen}. An example of WSI of liver cancer as well as tumoral region and normal region extracts are provided in Fig. 1.
However there are many variables during staining operation. The specimen thickness, concentration of the stain, manufacturer, time, and temperature can drastically change the overall appearance of the same tissue. 
The analysis of hispathological image focuses on the texture of cells, which is emphasized by the corresponding stain material. In the perspective of considering levels with lower resolution, as the cells have the characteristic of high-density regional aggregation, images tend to have regional color feature, as shown in Fig. 1. Therefore, color normalisation is required to homogenize the visual properties of the data.
\begin{figure*}[h]
	\centerline{\includegraphics[width=1.6\columnwidth]{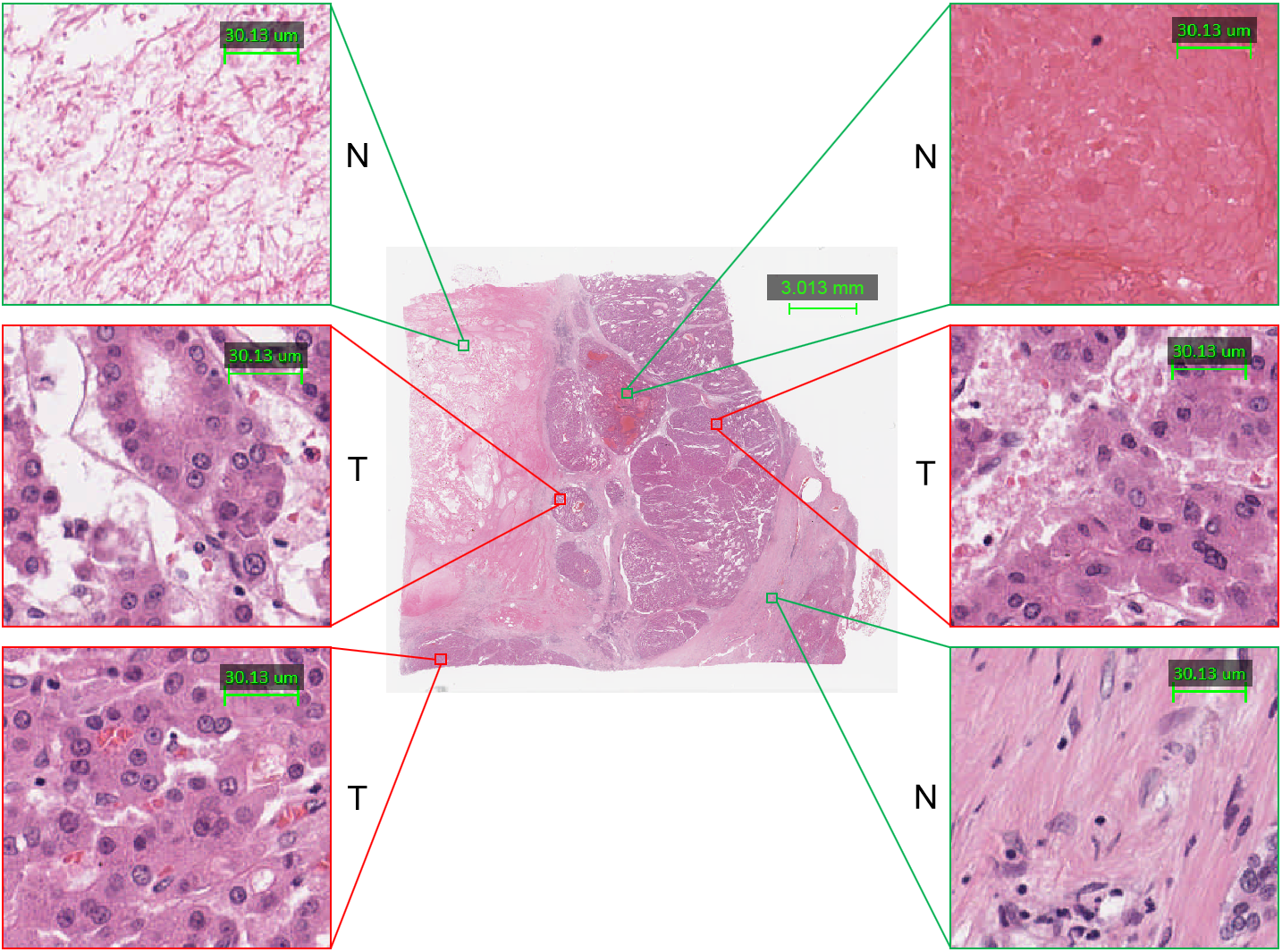}}
	\caption{An example of a whole slide image with 6 zoomed-in regions to illustrate the diversity of pathomorphology in tumor and normal tissues, and the complexity of interlaced foreground and background. Regions with green rectangles (with label N) are normal, with red rectangles (with label T) show tumor areas. }
	\label{fig}
\end{figure*}
Lastly, even if pyramid representation decreases the size of WSI, the size of one single downsampled image is still too large for acceptable computation. 
Therefore, weighted overlapping blocs are used in the forward procedure of dividing WSI into patches and backward procedure of assembling the patches back to WSI, which guarantes the completeness of WSI.

Nowadays, WSI serves as an enabling platform for the application of artificial intelligence (AI) in DP. Recently learning algorithms have emerged as a powerful approach for vision task. Due to their dominating performance in many tasks of computer vision, they have significantly impacted all the research areas in computer vision such as object classification, object detection and segmentation.
Since Krizhevsky et al. proposed AlexNet\cite{Krizhevsky2017ImageNet} and won the ImageNet Large-Scale Visual Recognition Challenge (ILSVCRC)\cite{ILSVCRC} as the first deep neural network algorithm, CNNs won these types of image analysis competitions consistently.
VGG(16-layer net)\cite{Simonyan2014Very}, GoogLeNet\cite{Szegedy2014Going}, ResNet\cite{He2016Deep} and DenseNet\cite{Huang2016Densely} et al., all of those CNNs have shown outstanding performances. The development of CNNs also provides a significant tool in the research field of medical imaging, such as radiography, magnetic resonance imaging, ultrasound, computer tomography\cite{Litjens2017A}.
DP as a relatively new area is an appropriate subject for the application of CNNs. Some researches focused on segmenting or detecting nuclei, epithelium, tubules, lymphocytes, mitosis, invasive ductal carcinoma and lymphoma directly\cite{Janowczyk2016, Liu2017Detecting, Cruzroa2014Automatic, 7780635}.
Some researches focused on using CNNs to obtain the pathological features of the specific diseases\cite{Qaiser2018Fast, Malon2013Classification, Le2012Learning}. However, the output of CNNs is block-based, this results in a too coarse segmentation for the task at hand.
 \begin{figure*}[htbp]
	\centerline{\includegraphics[width=1.3\columnwidth]{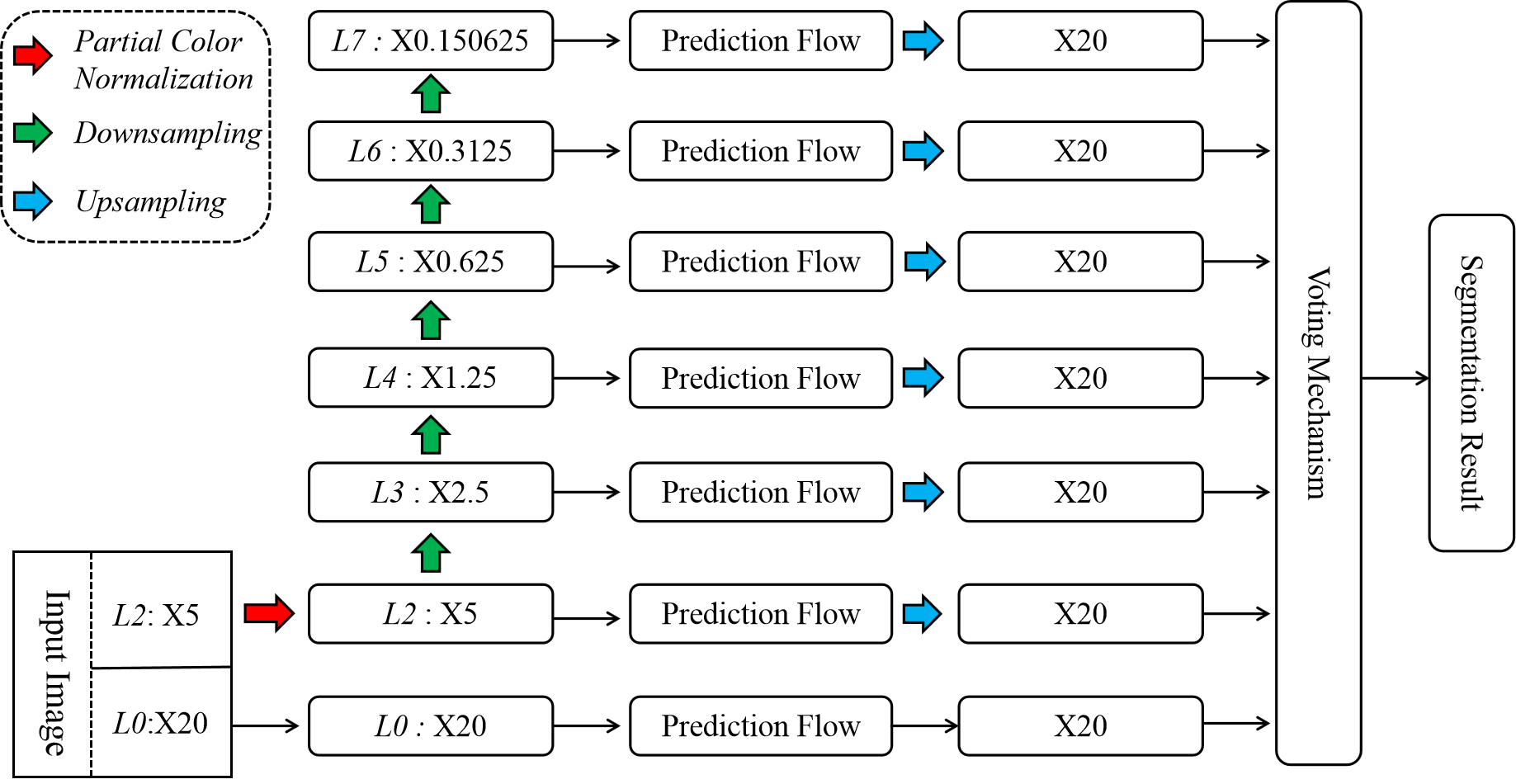}}
	\caption{Overview of the proposed method for cancer area segmentation in histology images. {$L_{n}$} corresponds to the {$n^{th}$} level of the pyramid representation. {$X_{r}$} corresponds to the image considered in the relative reduction case.}
	\label{fig}
\end{figure*}
 
With the emergence of Fully Convolutional Network (FCN)\cite{Long2015Fully}, the capability of networks has been explored to pixel-level labeling problem like semantic segmentation.
FCN transforms those existing and well-known classification models into fully convolutional ones by replacing the fully connected layers with convolutional layers and upsampling layers to output spatial maps instead of classification scores. Since FCN is composed of convolutional, pooling and
upsampling layers, depending on the definition of a loss function, it can be trained end-to-end.
After FCN, researchers explored more neural network architectures. Most of them adopt the encoder-decoder architecture. Encoder usually employs the architecture of classical CNNs to downsample the input image and abstract the features. Decoder upsamples the low-resolution maps outputed by encoder to the resolution of input image and achieves the pixel-wise classification.
Ronneberger et al.\cite{Ronneberger2015U} proposed U-Net to segment the neuronal structures in electron microscopy images. DeconvNet\cite{DBLP:journals/corr/NohHH15}, SegNet\cite{Vijay2017SegNet}, Pyramid scene parsing network (PSPNet)\cite{Zhao2016Pyramid} and DeepLab V1-V3\cite{Chen2017Rethinking, Chen2014Semantic, Chen2018DeepLab} have been proposed for scene segmentation firstly. Afterwards, they have been applied to the segmentation of medical images\cite{DBLP:journals/corr/abs-1802-02611, inproceedings}.  
Some networks use attention module. Oktay et al.\cite{Oktay2018Attention} proposed Attention U-Net to detect the pancreas. Sinha et al.\cite{DBLP:journals/corr/abs-1906-02849} proposed a multi-level attention based architecture for abdominal organ segmentation from MRI images. Some researchers focused on the segmentation of histopathological image\cite{Al2019Integrating,Xu2016A}.  
Some networks attempt to optimize the existing networks in the respects of reducting computation complexity and improving accuracy, such as Auto-DeepLab\cite{Liu2019Auto}, Global Convolutional Network (GCN)\cite{Peng2017Large} et al.. 
In this article, eight networks mentioned above are chosen to be compared in segmenting liver cancer area on pathological images. Different architectures show various performance in this task. Based on a quantified evaluation of the results of networks, the best one is used afterwards in the multiscale image processing method we propose for segmentation of liver cancer area. 

In this paper, we propose a new method based on deep learning segmentation with pyramidal decomposition approach.
The main contributions of this paper are as follows:
\begin{itemize}
	\item A multi-scale image processing method is proposed to specifically segment liver cancer area in Whole Silde Image (WSI). WSI is obtained at very high resolution, the pyramid representation provides perspectives in different scales. Through a voting mechanism, predictions in distinct levels are combined, which takes into account local and macro features at different scales. 
	\item Partial color normalization of tissue sections in the image was adopted to overcome the difference in staining procedure, which equalizes the color representation of different tissue areas.
	\item A weighted overlapping method is used to obtain the prediction of patch based network, which removes the seams between blocks. 
\end{itemize}

\section{Method}
\subsection{Dataset Description}
A tumor is composed of various cellular and stromal components, eg tumor cells, inflammatory cells, blood vessels, acellular matrix, tumor capsule, fluid, mucin, or necrosis.
The need for evaluation of viable tumor burden is increasing, as an assessment of response rates for chemoradiotherapy or proportion of tumor cells in genetic testing using tissue samples. 
The dataset used in this research comes from the 2019 MICCAI PAIP Challenge\cite{dataset}.
The original dataset contains 50 WSIs and ground truths for training, 10 WSIs and ground truths for validation, 40 WSIs for test.
All WSIs were stained by hematoxylin and eosin and scanned by Aperio AT2 at x20 power.
The goal of the challenge is to evaluate new and existing algorithms for automated detection of liver cancer in WSIs, focusing on the task of liver cancer segmentation. 

A large dataset is crucial for the performance of the deep learning model. Neural networks are expected to be robust in a variety of conditions, such as different orientation, location, scale, brightness etc. However, the datasets acquired in DP are small and taken in a limited set of conditions.  Data augmentation is then necessary,  based on the principle that the images created must be related to original images and as realistic as possible. According to the morphological characteristics of tissue sections and possible situations in the procedure of making WSI, data augmentation techniques used here are flip, rotation, crop and translation. They are systematically applied to all training datasets  considered in this paper. 
\begin{figure*}[htbp]
	\centerline{\includegraphics[width=1.6\columnwidth]{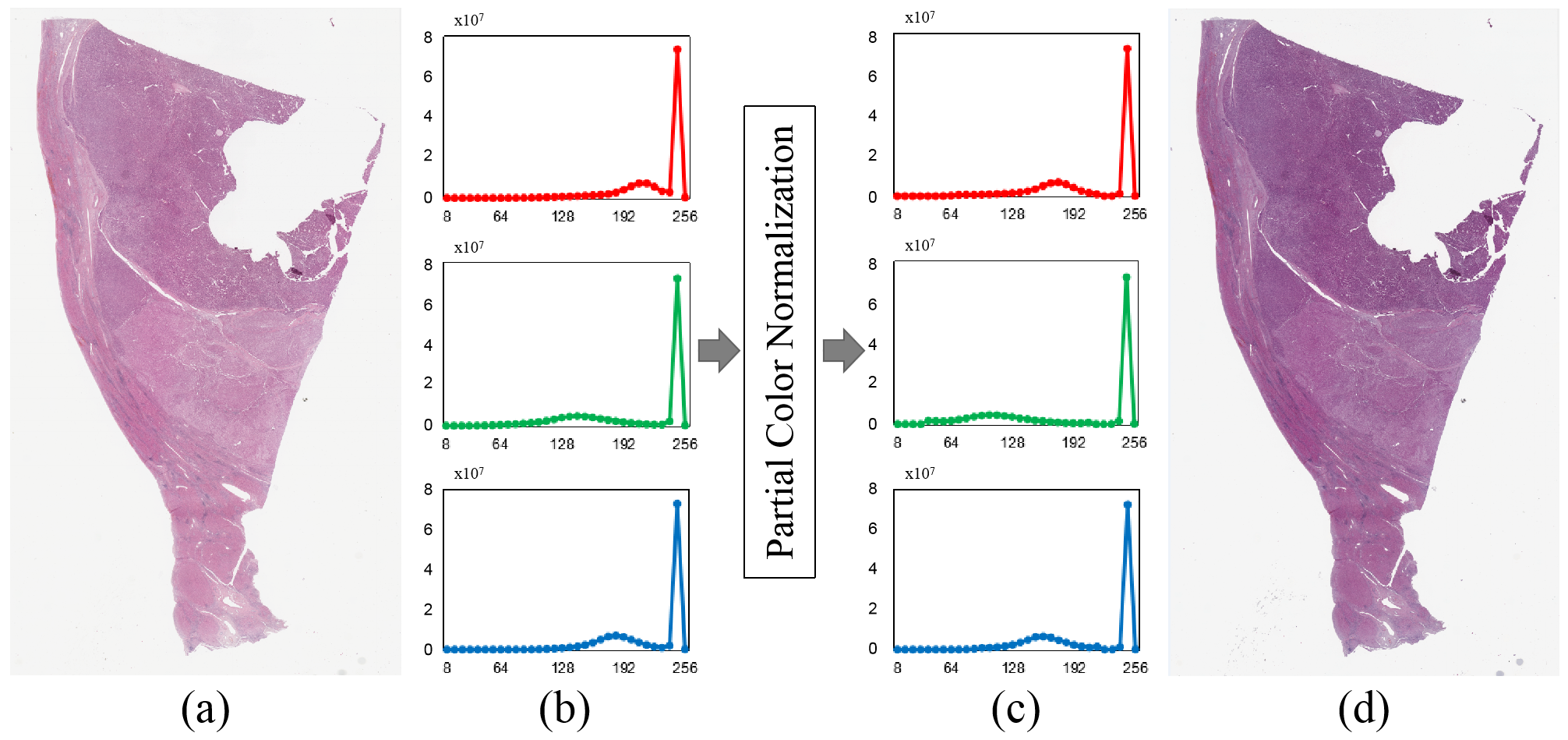}}
	\caption{Illustration of partial color normalization. (a) Original WSI of a liver cancer tissue section. (b) The histograms in R, G and B channels of original WSI. (c) The histograms in R, G and B channels of WSI after partial color normalization. (d) The WSI after partial color normalization.}
	\label{fig}
\end{figure*}
\subsection{Multi-scale image processing method}
An overview of the general procedure we propose for cancer area segmentation  is shown in Fig.2 while the prediction flow block is detailed in Fig. 4.
Firstly, FCN, DeconvNet, U-Net, Attention U-Net, Segnet, PSPNet, GCN and Deeplab V3 were trained with the same dataset. Those 8 networks are designed with unique network structures, which could result in variance when they are applied to solve the identical problem. The trained networks were evaluated on same dataset, and the results were quantified and compared. In the end, U-Net, which is the best one in the comparison, is chosen to be used within the multi-scale image processing approach.
Secondly, partial color normalization is used to correct the difference in the staining procedure before computing the pyramidal structure. Then, the gaussian pyramid representation is built for every WSI, which is broken down into 7 levels going from the original WSI's $0^{th}$ level to $7^{th}$ level. Each pixel in higher level is formed by the contribution from 5 pixels in underlying level with gaussian weights. By doing so, a M $\times$ N image at level \emph{i} becomes a M/2 $\times$ N/2 image at level \emph{i+1} and at each iteration areas consequently shrank to one-fourth. After that, 7 training datasets are constructed for each used level using the corresponding images. During this procedure, data augmentation is utilized to generate more images to train the network more robustly. Thirdly, the model of U-Net from the comparative study is taken as a pre-trained model, and this model is fine-tuned on 7 datasets to construct 7 models, one for each considered level. Fourthly, models are used to predict each pixel class and segment images in corresponding levels. The overlapping method is used here, which includes techniques about shifted cropping and weighted assembling. Fifthly, by a reverse pyramidal operation, the images from $2^{th}$ level to $7^{th}$ level are projected back to the same size of $0^{th}$ level, and the 7 predicted images are combined by a voting mechanism.

\begin{figure}[htbp]
	\centerline{\includegraphics[width=0.8\columnwidth]{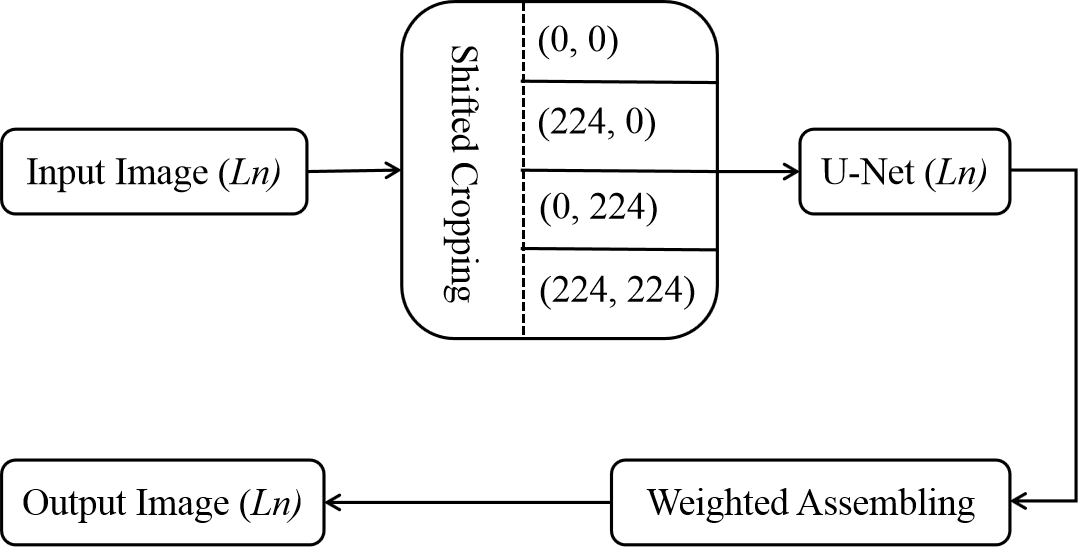}}
	\caption{Illustration of the prediction flow.}
	\label{fig}
\end{figure}

\subsection{Partial Color Normalization}
Histological slides are stained with multiple dyes to colour different types of tissues\cite{Ghaznavi2013Digital}. Colour consistency in light microscopy based histology is an increasingly important problem with the advent of Gigapixel digital slide scanners and automatic image analysis. In \cite{Reinhard2001Color}, Reinhard et al. presented a method for matching the color distribution of an image to a target image by using a linear transform in a perceptual colourspace (the $\l\alpha\beta$ colourspace of Ruderman et al.\cite{article}) so as to match the means and standard deviations of each colour channel in the two images in that colourspace. Wang et al.\cite{WangA} applied this method pixelwise on digital histology images stained with the Haemotoxin and Eosin (H\&E) stains. 

In the present study, WSI is scanned from the whole microslide, which means WSI contains a relatively large part of blank background. In addition, there are some areas where the tissue is too thick, and these areas mix with tissue of normal thickness, as can be observed in Fig. 1. The high-brightness background light affects the value of mean and deviation of the WSI and color normalization can stain the blank background area, which can mislead the segmentation result. In order to avoid this, we use partial color normalization as described in eq.1:
\begin{large}
	{\setlength\abovedisplayskip{12pt}
	\setlength\belowdisplayskip{6pt}
	\begin{equation}
	\left\{\begin{matrix}
	\l_{mp}=\frac{\l_{op}-\bar{\l}_{op}}{\hat{\l}_{op}}\hat{\l}_{tp} + \bar{\l}_{tp}
	\\
	\\ \alpha_{mp}=\frac{\alpha_{op}-\bar{\alpha}_{op}}{\hat{\alpha}_{op}}\hat{\alpha}_{tp} + \bar{\alpha}_{tp}
	\\
	\\ \beta_{mp}=\frac{\beta_{op}-\bar{\beta}_{op}}{\hat{\beta}_{op}}\hat{\beta}_{tp} + \bar{\beta}_{tp}	
	\end{matrix}\right.
	\end{equation}}
\end{large}

\noindent where $\bar{\l}$,  $\bar{\alpha}$ and  $\bar{\beta}$ are the channel means, $\hat{\l}$,  $\hat{\alpha}$ and  $\hat{\beta}$ are the channel standard deviations, the subscripts \emph{op}, \emph{mp} and \emph{tp} mean the foreground pixels in the original image, mapped image and target image respectively. The color normalization is operated in the $\l\alpha\beta$ colourspace, pixels from which the mean and deviation are calculated are limited in the tissue area using a mask. This mask is obtained by thresholding each WSI with the RGB value of (235, 210, 235), which is officially given by the challenge to separate the background and the holes within the tissue. Only within the mask pixels from the mapped image are transformed. Consequently, the effect of background on color normalization is excluded. Fig. 3 (b) and Fig. 3 (c) respectively present the histograms of R, G, B channels before and after partial color normalization. As can be seen, the histograms show three highest peaks. The pixels whose values belong to or around those three peaks are the background. Comparing the original histograms and the ones after color normalization, it can be seen that the pixels of background stay same and the pixels of tissue are tranformed. A result of partial color normalization is shown in Fig. 3 (d).

\subsection{Prediction}
The huge size of WSI is a big challenge in terms of computer memory and computation. Taking account of the cascading and accumulation calculation of the network during the train and prediction, it is impossible to use a whole WSI directly in standard computers. So, we have to crop the WSI to adapt it to the computer memory. But this operation brings another new problem for the semantic segmentation: the boundary area does not have enough information to support the neural network to make a precise decision. When we integrate the patches together, there are always seams at the boundary of blocks. To solve this problem, Cui et al.\cite{Cui2018A} proposed a method of overlapped patch extraction and assembling in the task of segmenting the nuclei on histopathology images. Here we refined the method in terms of weight map calculation and applied it only in the procedure of assembling.

\begin{figure}[htbp]
	\centerline{\includegraphics[width=1\columnwidth]{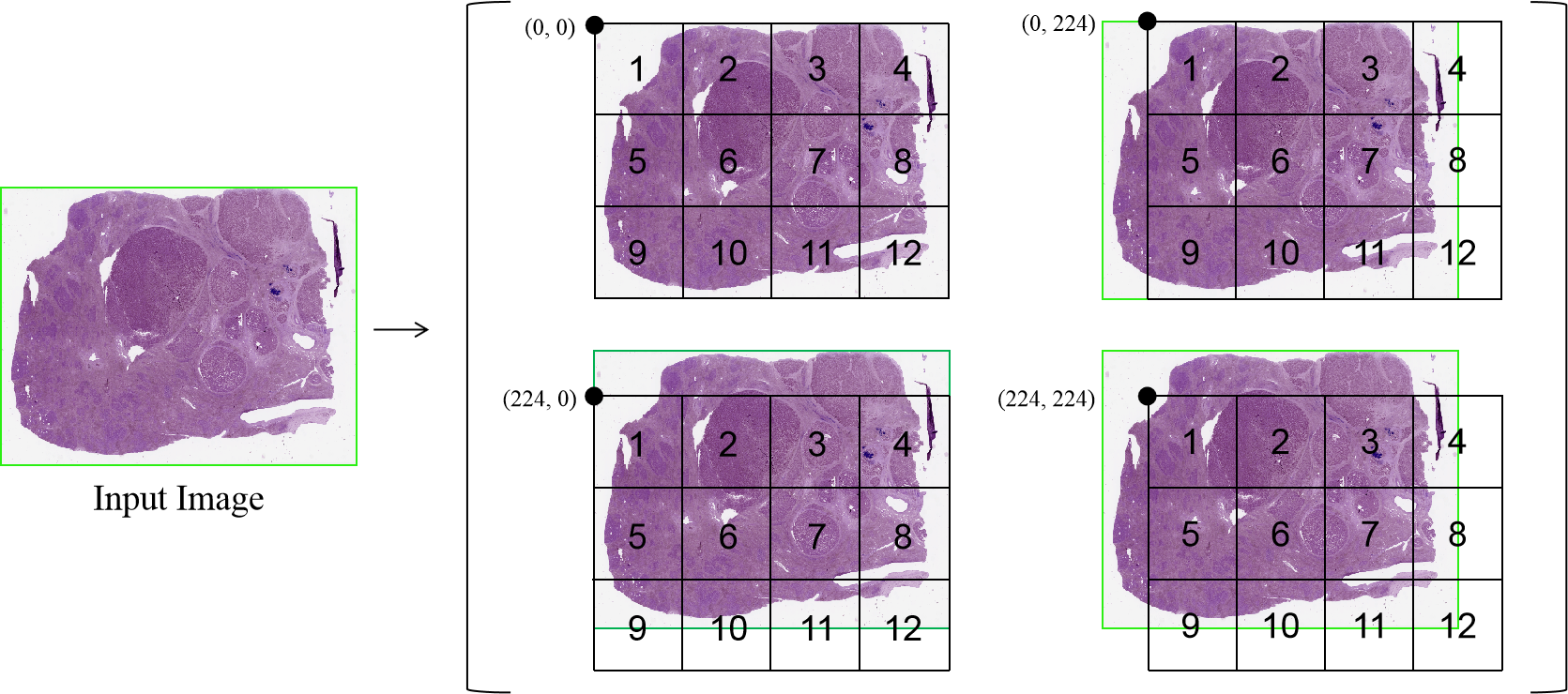}}
	\caption{Illustration of the method of shifted cropping. The original image is cropped into four groups with the shift (0, 0), (224, 0), (0, 224) and (224, 224). The size of each cropped patch is 448x448 pixels.}
	\label{fig}
\end{figure}

The input image of each level $L_{n}$ is processed by the shifted cropping firstly, which is demonstrated in Fig. 5. In this work, the size of input patch is 448x448 pixels, shifted distance here is taken as half of the patch size. The cropping operation is composed of none-shift (0,0), x-shift (224,0), y-shift (0,224) and x-y-shift (224,224). Then the input image is divided into 12 patches in each group. After that, patches are sent into the U-Net by group separately. 

The four groups of predicted images are combined to the final image in the weighted assembling procedure, which is described in eq.2:
\begin{large}
	{\setlength\abovedisplayskip{12pt}
	\setlength\belowdisplayskip{10pt}
	\begin{equation}
	F(i,j)=\frac{\sum_{k}W({P_{k}(i,j)})\cdot G^{k}(P_{k}(i,j))}{\sum_{k}W({P_{k}(i,j)})}
	\end{equation}}
\end{large}
\noindent where \emph{F(i, j)} is the pixel at position (i, j) on final predicted image. \emph{$P_{k}(i,j)$} is the coordinate transformation for mapping (i, j) to \emph{$k^{th}$} group. \emph{$G^{k}$} is predicted map of \emph{$k^{th}$} group. \emph{$W$} is weight map as described in eq.3:
\begin{large}
	{\setlength\abovedisplayskip{12pt}
	\setlength\belowdisplayskip{12pt}
	\begin{equation}
	\left\{\begin{matrix}
	W(i,j)=\frac{D_{i,j}^{b}}{D_{i,j}^{b}+D_{i,j}^{c}} \\
	\\
	D_{i,j}^{b}=min(D_{i,j}^{b_{1^{th}}},D_{i,j}^{b_{2^{th}}},D_{i,j}^{b_{3^{th}}},D_{i,j}^{b_{4^{th}}})
	\end{matrix}\right.
	\end{equation}}
\end{large}
\noindent where \emph{$W(i, j)$} is the weight of pixel (i, j); \emph{$D_{i,j}^{c}$} is the \emph{$L^{2}$} distance of pixel (i, j) from the center pixel; \emph{$D_{i,j}^{b}$} is the perpendicular distance from border, the shortest one of pixel (i, j) from the four borders is taken.

\begin{figure}[h]
	\centerline{\includegraphics[width=0.9\columnwidth]{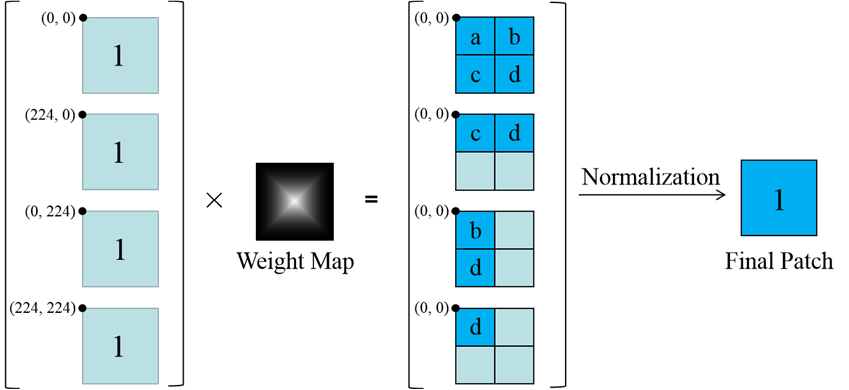}}
	\caption{Weighted assembling procedure. The left column contains the four $1^{th}$ patches from shifted cropping in Fig. 5. The right column contains the patches after multiplying by weight map. The areas labeled with \emph{a, b, c, d} are mapped to contribute to the final patch, the blocks with same label are overlapped.} 
	\label{fig}
\end{figure}
Since the shifted cropping, the shifted images lose the boundary pixels of original image. At the same time, the block pixels are multiplied by the weight map in the reconstructed pixel coordinates. 
As shown in Fig. 6, the four first patches at the corner of image, which loses pixels in three direction respectively, are extracted to explain the weighted assembling procedure and boundary processing method.
The final patch is constructed by four patches, blue blocks are the area where the original image can be mapped to.
Only the blue areas with same letter can be overlapped. For the area lost during shifted cropping (for example the block \emph{a} is lost in the other three groups), the boundary processing method of eq.2 sets the pixels and weights of missing area to 0.

\begin{figure*}[ht]
	\centerline{\includegraphics[width=2\columnwidth]{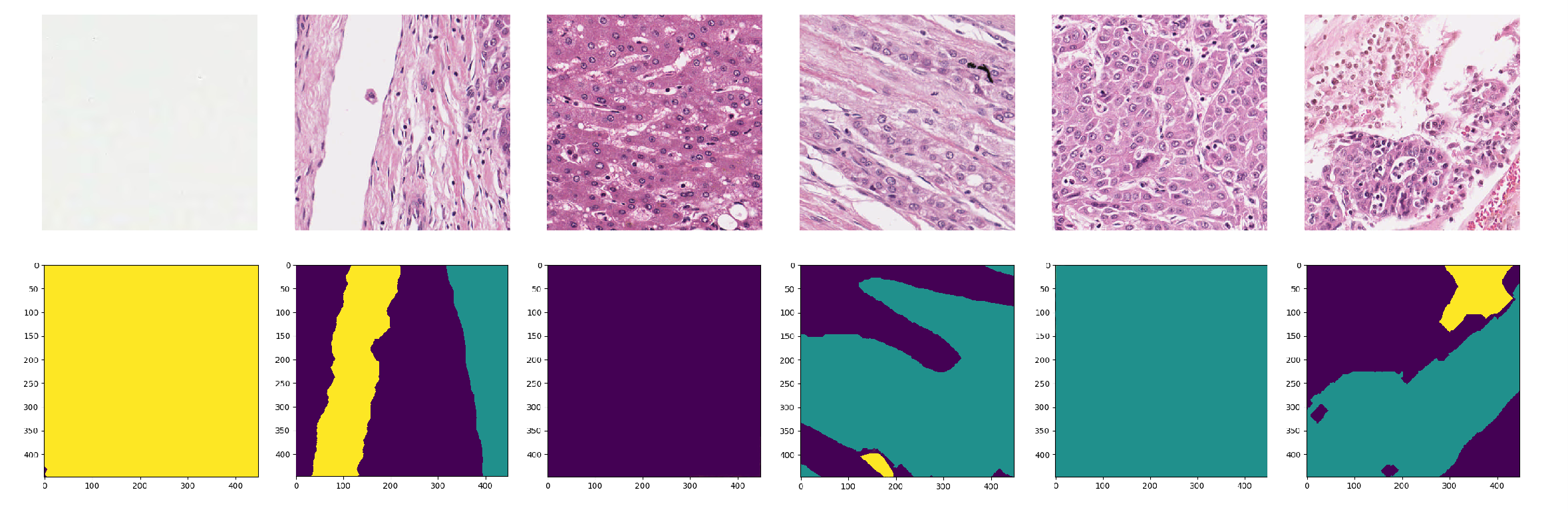}}
	\caption{Examples of original histopathology images in first row, the corresponding ground truths in second row. Within the ground truth images, yellow, black and blue areas represent background, normal tissue and tumor tissue respectively.}
	\label{fig}
\end{figure*}

\subsection{Combination}
In the proposed architecture, we build during the downsampling process a pyramidal representation of the WSI (L0 to L7) to obtain multi-scale features. For the combination of the 7 obtained predicted images, we employ a voting mechanism to obtain the final result. Specifically, we use Gaussian pyramid operation to project the images from the top layers back to the bottom level. This upsampling step allows having images of same size for all considered levels before applying the voting process. It should be noticed that the predicted images are probality maps. As the value of each pixel is between 0 and 1, we used a threshold \emph{(Th)} to classify a given pixel. Then, as described in eq.4, if the pixel at position (i, j) is valid in more than \emph{N} layers, it is considered as \emph{TRUE} in the final segmentation result:
\begin{large}
	{\setlength\abovedisplayskip{12pt}
	\setlength\belowdisplayskip{12pt}
	\begin{equation}
	V(i,j)=\left\{\begin{matrix}
	1,& if\; \sum_{k=0}^{7}v_{k}(i,j)>=N& \\ 
	& &\\
	0,& if\;  \sum_{k=0}^{7}v_{k}(i,j)<N&
	\end{matrix}\right.\label{eq}
	\end{equation}}
\end{large}

\noindent where \emph{V(i, j)} is the final prediction of the pixel at position (i, j) in an image. \emph{$v_{k}(i, j)$} corresponds to the prediction in the \emph{$k^{th}$} layer. \emph{N} is the chosen threshold that controls the majority voting mechanism.  

\section{Experiment}
\subsection{Evaluation of deep segmentation architectures}
The dataset used in the experiment was constructed from the original 50 WSIs of training. 
The images in the dataset were cropped randomly from the $0^{th}$ level of WSI. 
The size of extracted patches is of 448 x 448 pixels, the dataset is built with 100,000 images. 
We set 20\% of this training set aside for evaluating the performance of eight models, namely: FCN, DeconvNet, U-Net, Attention U-Net, Segnet, PSPNet, GCN and Deeplab V3. 
In addition, the label of ground truth is modified to three sorts, with background tags added. 
The original dataset doesn't provide the label of background area, the label used here is obtained by the thresholding procedure provided by the challenge as indicated in section \uppercase\expandafter{\romannumeral2}. C. Fig. 7 presents examples of the constructed dataset. 
The input of networks is set as 448 x 448 pixels with RGB channels, the output of networks is an image of 448 x 448 pixels with tumor tissue, normal tissue and background area channels. 
As the amount of parameters to be trained varies significantly with each network, the training time for each network is not unified. 
So, the final trained model is chosen when the difference of the \emph{validation accuracy} of each epoch stays within the tolerance of 0.001 ten times during the training.  

\begin{table}[htbp]
	\centering
	\caption{The evaluation (S: score; R: rank) result of networks}
	\begin{tabular}{llrrlrrlr} 
		\toprule           
		&\multicolumn{2}{c}{F1 score} &             & \multicolumn{2}{c}{Jaccard score} &               & \multicolumn{2}{c}{Hausdorff d.}  \\
		\cmidrule{2-3}                              \cmidrule{5-6}                                      \cmidrule{8-9}
		&S      &R                    &             & S      & R                  		&               & S      & R                       \\
		\midrule
		U-Net        		&\textbf{0.465}  &\textbf{1}              &  		&0.904        &2          & 				&\textbf{4.793}         &\textbf{1}                  \\
		Deeplab V3   		&0.444  &3              & 	    &\textbf{0.906}        &\textbf{1}          &  				&5.003         &3                  \\
		GCN  		        &0.442  &4              & 		&0.891        &3          &  				&4.919         &2                   \\
		SegNet       		&0.445  &2              &  		&0.813        &5          &  				&8.017         &5                   \\
		DeconvNet    		&0.441  &5              &  		&0.819        &4          & 				&6.541         &4                  \\
		PSPNet       		&0.42   &7              &  		&0.742        &6          &  				&8.436         &5                   \\
		Atten-UNet   		&0.432  &6              &  		&0.633        &7          &  				&9.292         &7                  \\	
		FCN          	    &None   &8              & 	    &None         &8          &  				&None          &8                  \\
		\bottomrule
	\end{tabular}
\end{table}

\begin{figure*}[h]
	\centerline{\includegraphics[width=2.1\columnwidth]{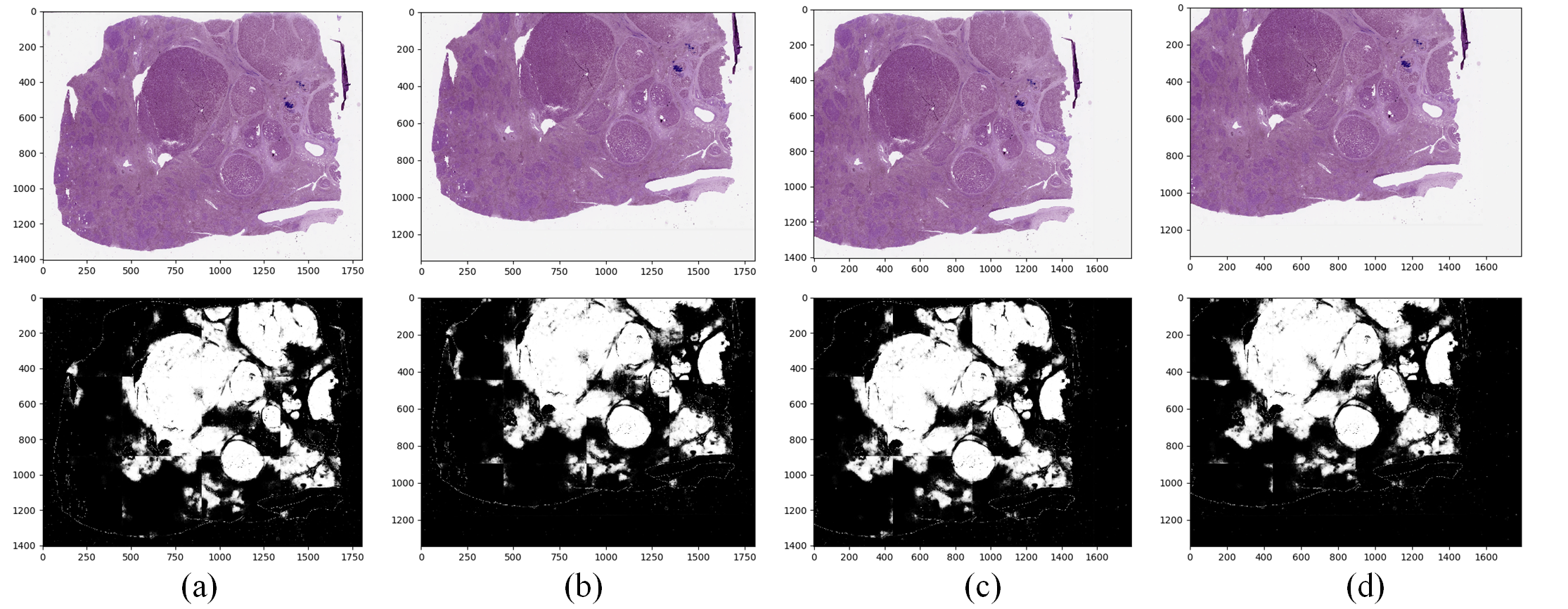}}
	\caption{Example of shifted cropping images in $5^{th}$ level and corresponding predicted results. (a) is shifted (0, 0) in x-y direction, so it can be regarded as the result of regular block-based segmentation method. (b) is shifted (224, 0) in x-y direction. (c) is shifted (0, 224) in x-y direction. (d) is shifted (224, 224) in x-y direction.}
	\label{fig}
\end{figure*}

The quantitative comparison is listed in Table \uppercase\expandafter{\romannumeral1}. The performance of networks is assessed by evaluation criteria consisting of F1 score, Jaccard similarity score and directed Hausdorff distance\cite{Taha2015An}. It can be observed that U-Net achieves the best performance compared to the other networks although it doesn't obtain the best one for the Jaccard score, but the difference is relatively small. The FCN gets the worst performance and the results of evaluation are \emph{None}, the reason is that serious overfitting occurred in the training phase.

\subsection{Implementation of the pyramidal representation of WSI}
The richness of information in WSI provides pathologist enough histopathological characteristics for diagnosis. Meanwhile, the huge dimension of WSI is a big challenge for computer memory and computation generally. The pyramidal representation and related downsampling allow to reduce the computational cost at each level. However, the images in some levels are still too large to be computed by GPU in original format. On the other hand, to reuse the U-Net architecture and inherit the trained parameters in different levels, the images are cropped into 448 x 448 pixels patches in $0^{th}$, $2^{th}$, $3^{th}$, $4^{th}$, $5^{th}$, $6^{th}$ level and resized into 448 x 448 pixels in $7^{th}$ level. Fig. 8 shows an example from $5^{th}$ level of pyramid representation. The original size of this WSI is 44984 x 57767 pixels. After 5 times gaussian pyramid operation, the size is reduced to 1406 x 1806 pixels. 

\begin{figure}[htbp]
	\centerline{\includegraphics[width=1\columnwidth]{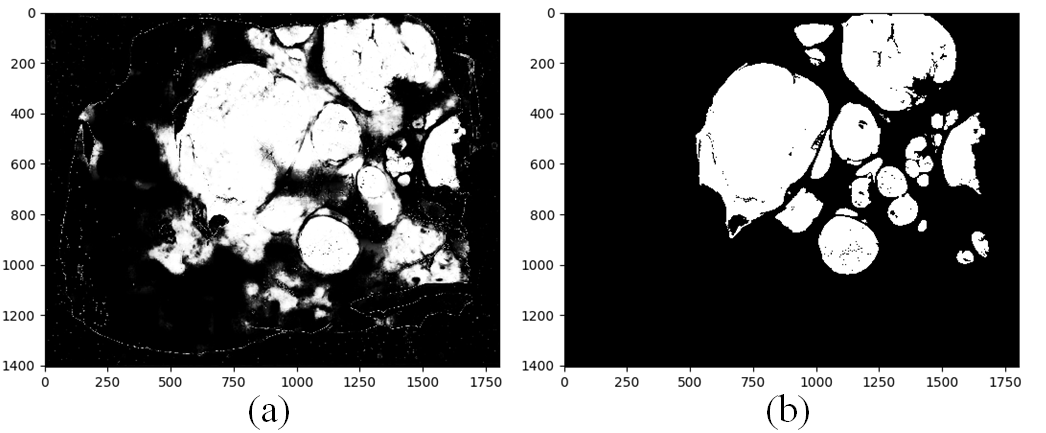}}
	\caption{(a) Weighted assembling result corresponding to the shifted cropping procedure depicted in Fig. 5 and Fig. 6 applied to the four images of Fig. 8 (a)-(d). (b) Related ground truth.}
	\label{fig}
\end{figure}

As shown in Fig. 8 (a), the image is cropped and predicted patch by patch in regular way. It can be seen that there exist obvious artifacts between each patch, and the predicted areas in adjacent patches do not match well. As described in \emph{Method} section, the other three shifted cropped images are computed in a complementary manner. They are shown in Fig. 8 (b)-(d). Identical problems to those observed on Fig. 8 (a) are generated in these three images. Then, these four images are combined using the weighted assembling method, the final result is shown in Fig. 9 (a). 

\begin{figure}[htbp]
	\centerline{\includegraphics[width=0.76\columnwidth]{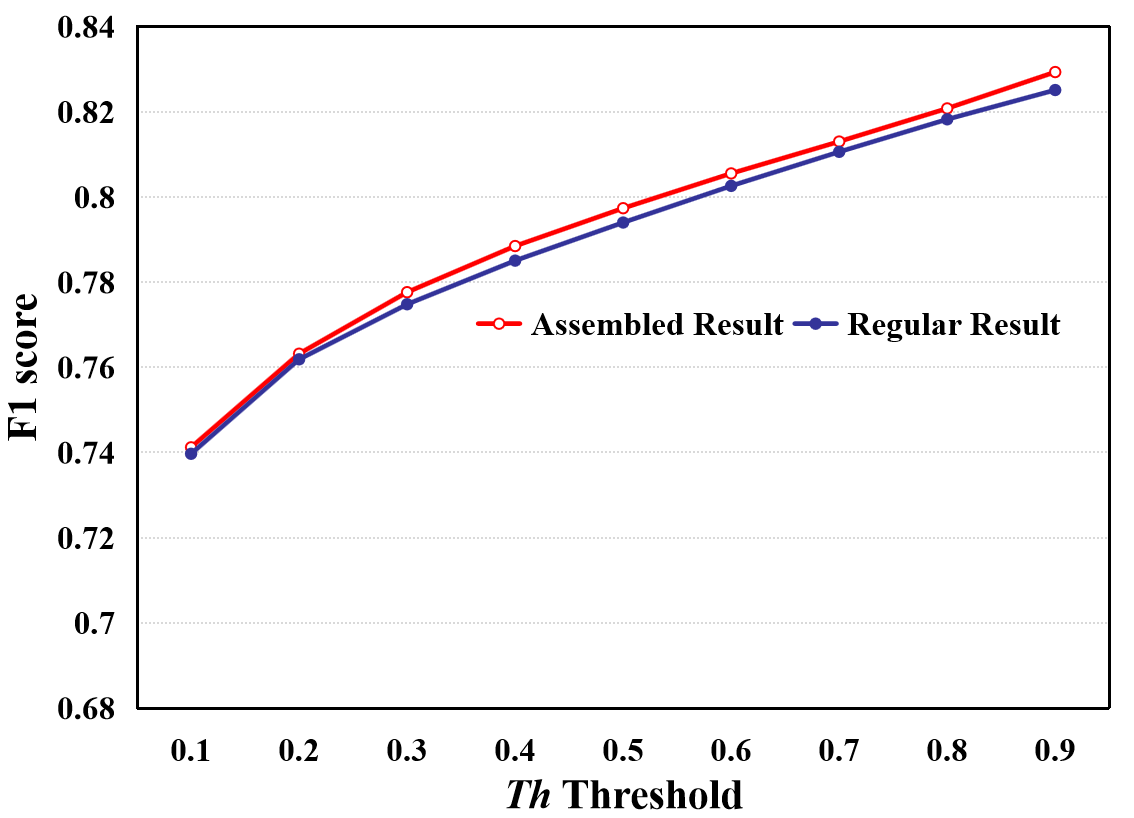}}
	\caption{Evaluation result of weighted assembling and regular block-based predicted results corresponding respectively to the images presented in Fig. 9 (a) and Fig. 8 (a).}
	\label{fig}
\end{figure}

\begin{figure*}[h]
	\centerline{\includegraphics[width=2\columnwidth]{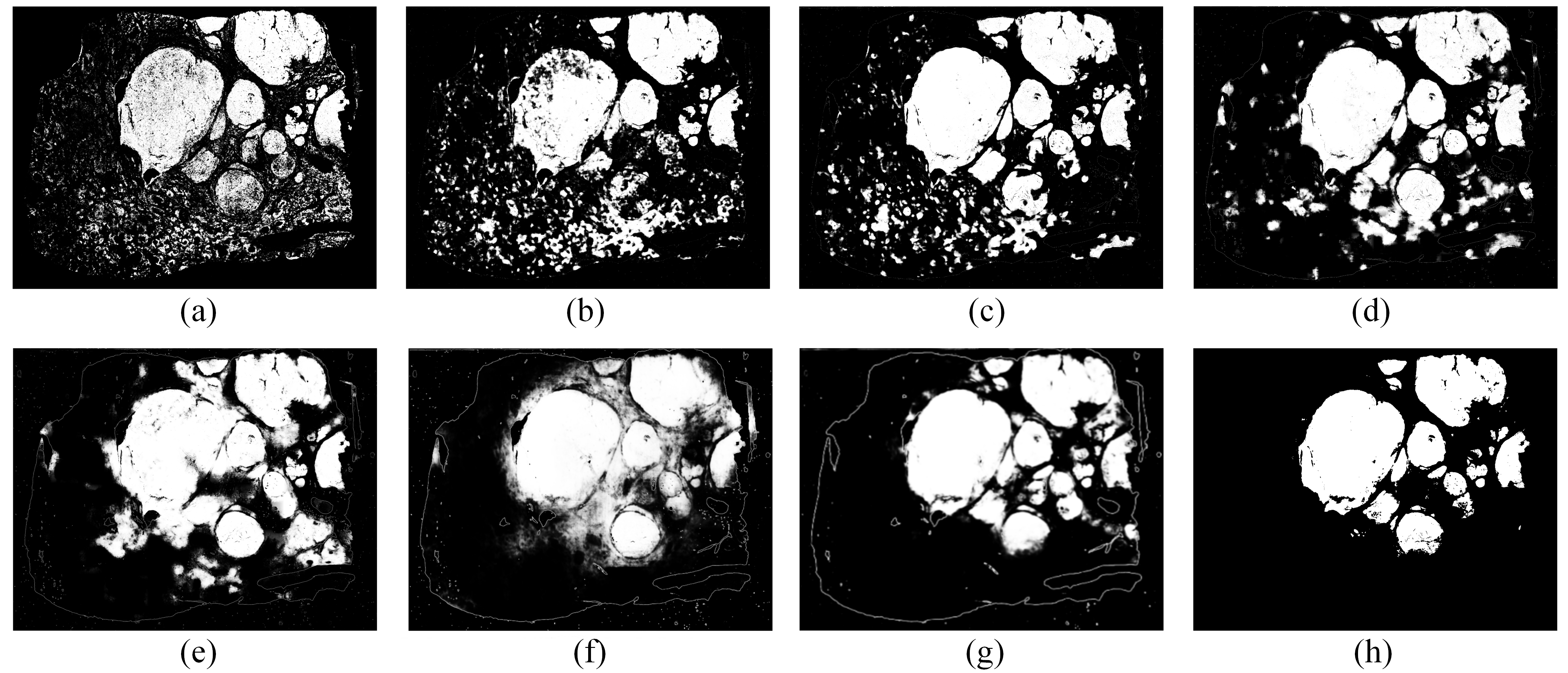}}
	\caption{The predicted results of 7 levels of the proposed pyramidal approach and result of voting mechanism. (a), (b), (c), (d), (e), (f) and (g) are the results from $0^{th}$, $2^{th}$, $3^{th}$, $4^{th}$, $5^{th}$, $6^{th}$ and $7^{th}$ level respectively. (h) is the result of voting mechanism. Threshold \emph{Th} is set at 0.7, majority vote number \emph{N} is set at 6.}
	\label{fig}
\end{figure*}

Through comparing the final result, shifted cropping results and original image, the final result effectively solves the problem of block matching. In addition, the final result mainly preserves correct predicted information from the four images, filters out errors and comes close to the ground truth shown in Fig. 9 (b), which is presented here using only two classes: tumor tissue and other according to the representation code used in the 2019 MICCAI PAIP Challenge. Fig. 10 presents the evaluations, with F1 score, of the weighted assembling result and the regular result, which corresponds to the block-based segmented result with (0, 0) shifted cropping presented in Fig.8 (a). Because the predicted results are probability maps, nine threshold \emph{(Th)} values are set from 0.1 to 0.9 in steps of 0.1 for the binarization. The F1 score shows that the predicted result gets promoted by the weigthed assembling procedure.

\subsection{Evaluation of voting mechanism}
This procedure is implemented at each level of the proposed pyramidal approach, which improves the integrity and accuracy of the predicted images. Then through upsampling operation, these predicted results are mapped to the original size of WSI. Compared with the ground truth presented in Fig. 9 (b), the predicted results of each level, presented in Fig. 11 (a)-(g), contain a large amount of errors. Since the directly predicted images are probalibility maps, a binarizition is conducted with threshold \emph{Th}. 
During the evaluation of images presented in Fig. 11 (a)-(g), the threshold value \emph{Th} serves as a variable to binarize the predicted probability maps. Table II presents the best results obtained for each level according to F1 score and Jaccard score. Then, each pixel with the value of 1 after binarization represents 1 vote, the binary images are used to vote for the final result in pixel level, presented in Fig. 11 (h) considering \emph{Th=0.7} and \emph{N=6}. The Jaccard score obtained for this segmentation result is 0.844 which corresponds to the best result compared with the directly predicted ones.
\begin{table}[htbp]
	\centering
	\caption{The evaluation of images in different levels.}
	\begin{tabular}{clcclcc} 
		\toprule           
					&\multicolumn{2}{c}{F1 score} &             & \multicolumn{2}{c}{Jaccard score} &               \\
					\cmidrule{2-3}                              \cmidrule{5-6} 
		&Score      &Threshold                    &             & Score      &Threshold                  		&           \\
		\midrule
		Fig. 9 (a)    		&0.834  &0.6                     &  			&0.575        &0.4                   		& 			\\
		Fig. 9 (b)       	&0.774  &0.6                     &  			&0.531        &0.6                    		& 		    \\
		Fig. 9 (c)   		&0.866  &0.9                     &  			&0.616        &0.8                    		&  			\\
		Fig. 9 (d)     	    &0.866  &0.8                     &  			&0.620        &0.9                    		&  			\\
		Fig. 9 (e)     	    &0.828  &0.9                     &  			&0.654        &0.9                    		&  			\\
		Fig. 9 (f)  		&0.838  &0.8                     &  			&0.746        &0.8                    		&  			\\
		Fig. 9 (g)   		&0.906  &0.6                     &  			&0.805        &0.8                    		&  			\\
		\bottomrule
	\end{tabular}
\end{table} 

Among the directly predicted images, the $6^{th}$ and $7^{th}$ layers outperform the others. We therefore changed the voting mechanism presented in eq.4 by according to $6^{th}$ and $7^{th}$ layers two voting rights, whilst every other layer has one voting right. So the total number of vote is finally 9 and the number of vote \emph{N} is another variable in the evaluation. Fig. 12 presents an expanded evaluation of the final segmentation result obtained after new voting mechanism. 
It can be seen that both of \emph{N} and \emph{Th} are related to the evaluation result. On the whole, when \emph{$N \geq 5$} and \emph{$Th \geq 0.5$}, the evaluation results have remarkable change, which means the intermediate point of the global threshold is crucial. All of the four curves prensented in Fig. 12 show the same trend: as a first step the evaluations raise with the increase of \emph{Th}. Meanwhile, the increase of \emph{N} also improves the outcome of the assessment. One can also notice that, when \emph{$N \geq 6$}, after reaching an optimum, the evaluation value significantly falls when the \emph{Th} threshold increases.

\begin{figure}[htbp]
	\centerline{\includegraphics[width=0.76\columnwidth]{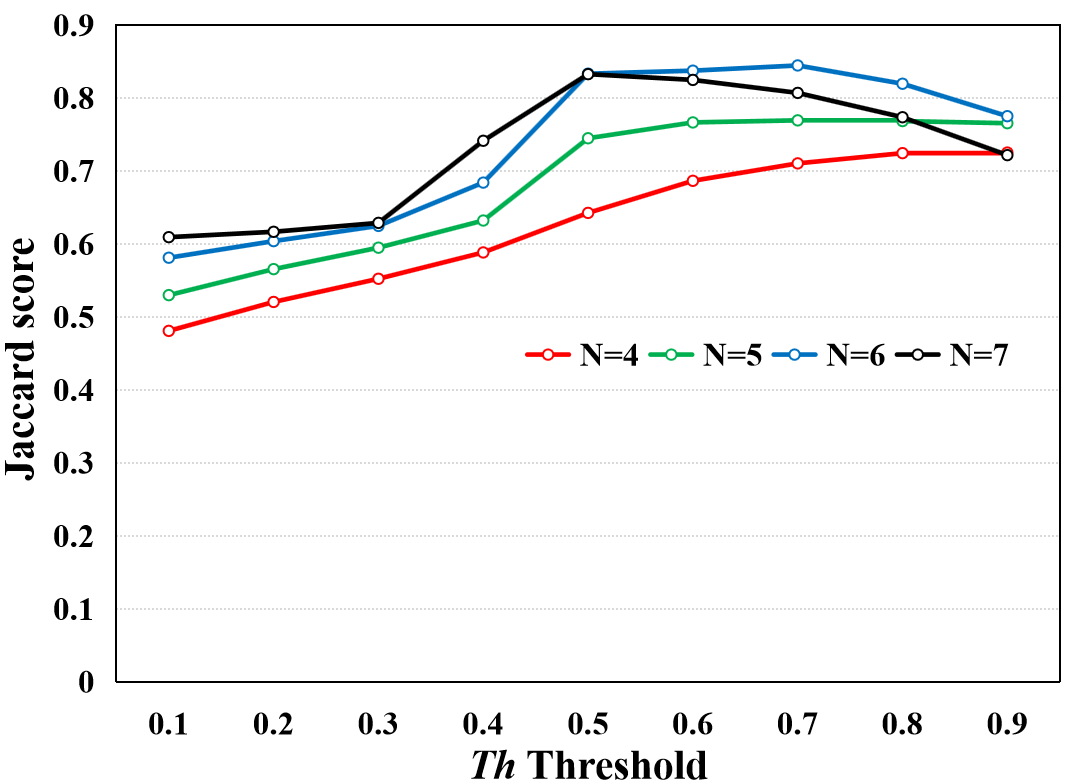}}
	\caption{Evaluation result of the voting mechanism corresponding to Fig. 11(a)-(g) with the parameters \emph{N} and \emph{Th}.}
	\label{fig}
\end{figure}

We finally applied this method to the other ten images provided by the 2019 MICCAI PAIP Challenge for validation. 
The results are shown in Fig. 14. Each curve in the left column corresponds to one validation image. 
It can be seen that the impact of \emph{Th} threshold on the final result shows a similar trend namely the quality of image will be enhanced after 0.5. 
The vote number \emph{N} is another factor. 
In this work, \emph{N=6} allows to achieve the best evaluation result in terms of mean score. 
Table \uppercase\expandafter{\romannumeral3} summarizes the results obtained by the 5 best ranked participants in the 2019 MICCAI PAIP Challenge\cite{challenge}. 

\begin{table}[htbp]
	\centering
	\caption{The comparison with 2019 MICCAI PAIP Challenge participants.}
	\begin{tabular}{clcclcc} 
		\toprule           
		&\multicolumn{2}{c}{Jaccard score} &                           \\
		\cmidrule{2-3}                              
		&S      &R                   			 &                 \\
		\midrule
		newhyun00    	&0.7298  &1                     &  				   \\
		ed15b020      	&0.7042  &2                     &  			 	   \\
		majianqiang  	&0.6995  &3                     &  			 	   \\
		gcggcg5     	&0.6975  &4                     &  		     	   \\
		bmarami     	&0.6614  &5                     &  			 	   \\
		ours  			&\textbf{0.7964}  &                   &  				   \\
		\bottomrule
	\end{tabular}
\end{table}

\subsection{Discussion}
\subsubsection{neural networks}
Different networks have been constructed with specific architectures. The textural or colour features that each network prefers differ from one another. Following, U-Net gets best result in the evaluation through F1 score and Directed Hausdorff distance, and second one in Jaccard score. It shows the stability in this comparison. In addition, another factor is also taken into account, that is the amount of parameters of networks. We compare in Fig. 13 the total parameters, trainable parameters and non-trainable parameters. It shows that U-Net also has the middle amount of parameters, which means faster training. Lastly, FCN is labelled in Table \uppercase\expandafter{\romannumeral1}, as \emph{None} in the results of evaluation. The reason is that FCN has serious overfitting. We also tested FCN in the WSI of breast tissue to segment the cell. In this case, it did not have overfitting. The reason for overfitting in segmentation of liver cancer could be that the object to be segmented presents fragmentization, so that the feature can not be obtained by FCN easily.

\begin{figure}[htpb]
	\centerline{\includegraphics[width=0.9\columnwidth]{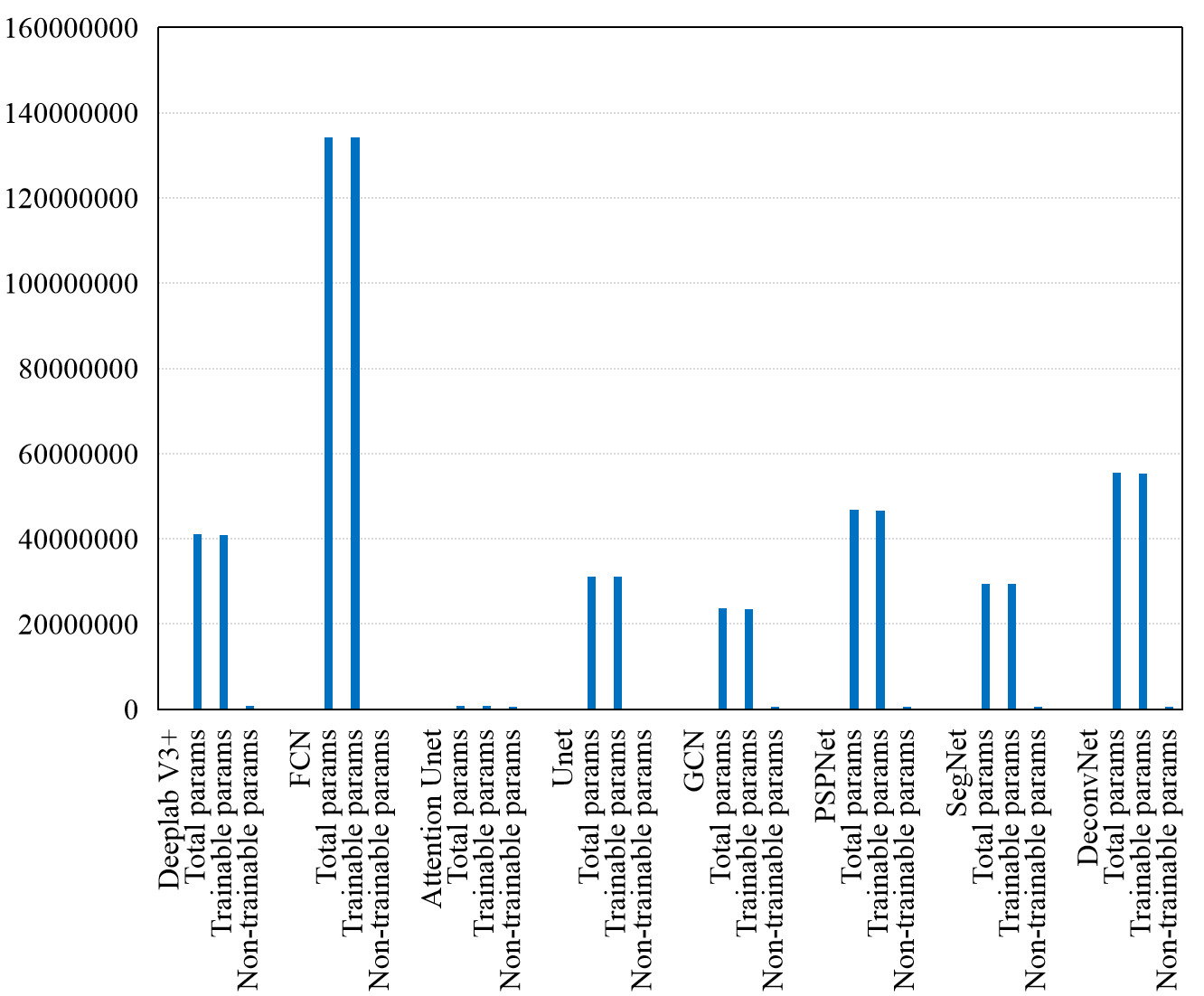}}
	\caption{Comparison of networks parameters.}
\end{figure}

\begin{figure*}[h]
	\centerline{\includegraphics[width=2\columnwidth]{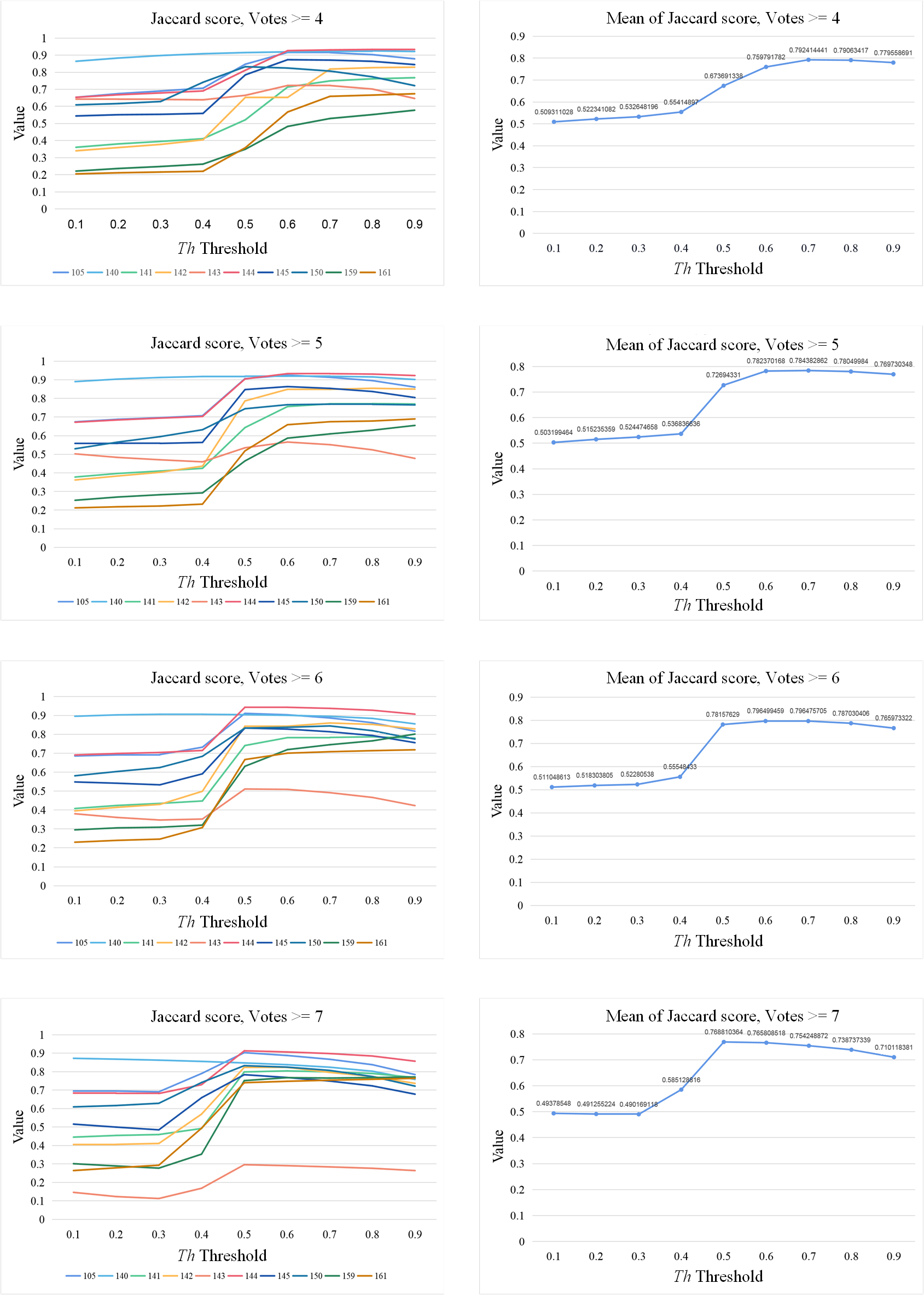}}
	\caption{The results of evaluation for the validations.}
	\label{fig}
\end{figure*}

\subsubsection{color normalization}
Color consistency is usely requested by CAD systems. 
In this instance, it is all the more important. 
What we propose is a multiscale image processing method and the pyramid representation is employed to construct different levels. 
The images of higher levels consequently lose the detailed features, such as pleomorphism, crowding, polarity, mitosis and nucleoli, which are crucial information used by pathologist in diagnosis. 
However, higher levels obtain larger view, specially the images of middle levels combine the features in detail and in larger view. 
Because the hematoxylin and eosin have different dyeing tendencies, the aggregation distribution of cells results in a regional color consistency on the images of higher levels.
As the size of the images gets smaller, the characteristics of the color area gradually emerge, which roughly indicates the area of the cancer cell. From the point of view of possible understanding of neural networks, color characteristics may be the main basis of neural network segmentation of pathological images in higher levels. 
To eliminate different variables during WSI's acquisition, the color normalization is necessary in this case as a significant preprocessing method for neural networking.

We adapted a partial color normalization based on the linear transform, which only concentrates on cellular tissue region. Actually, the regional white background represents a large component on WSI. Meanwhile, there are a lot of holes or very thin areas in the tissue, which belong to the background class too. For the network, background doesn't need to be normalized. But in the linear transform, if background area is taken into account, the background area will be stained and the color of tissue area will become shallow consequently. Moreover, the degree of change will depend on the ratio of background and tissue areas which is an uncertain variable. Partial color normalization allows to avoid the impact of white space and unifies the color effectively, preparing thus the images for the next processing steps. 

\subsubsection{shifted cropping and weighted overlap}
In the prediction part, shifted cropping and weighted overlap are used to produce a continuously complete image. Since WSI has a huge size, it is nearly impossible to input WSI into networks; therefore, cropping is neccessary. Meanwhile, cropping generates another problem: the continuity and integrity of image information are destroyed. In addition, the edge region of image can not provide enough information for neural network to do semantic segmentation preciselyand cropping creates patches which means more edge regions. Hence, it can be seen in the rawly assembled image that the predicted results are not complete. In this work, under the premise that the size of cropped patch is (448, 448), the image to be predicted is cropped from locations (0, 0), (0, 224), (224, 0), (224, 224), which means that the horizontal, vertical and cross boundaries are covered by the centre of blocks in corresponding overlapping layer. In the weighted overlapping procedure, the weight matrix has two functions, one is to weaken the border's predicted result of each block, another one is to emphasize the central information of each block. Shifted cropping and weighted overlap guarantee that the final predcited result is constructed by the central area of each block,  removing consequently the seams. 

\subsubsection{voting mechanism}
Voting mechanism is proposed in this work to combine the predicted results in different scales. Seven levels of gaussian pyramid representation are built, the size of image in each level is reduced to one-fourth of previous level. The WSIs of dataset are given in \emph{.svs} format, each file originally contains 3 or 4 layers of pyramid representation, but the downsample factor is 4. Taking account of the huge size of the $0^{th}$ level and the preservation of original data, the consecutive pyramid representation is generated from the original $2^{th}$ level with downsample factor of 2. Then, the U-Net is trained and predicts the image in each level independently. The U-Net model of $0^{th}$ level is trained with original images, the dataset for training is built using the data augmentation without color normalization, which provides the primary source of information. 

The predicted results of different layers are binarized firstly to finish the transformation from probability map to binary image. The best value of \emph{Th} threshold is an unknown quantity. The mean evaluation curves in this study (see Fig. 14.) show a trend of first increasing and then decreasing, which indicates that the best value of \emph{Th} is between 0.5 and 1. The selection of optimal threshold for specific problems needs further study. The binarization of an image also gives each pixel right to vote when its value is 1.
The result of voting mechanism takes the value when voting number is greater than the chosen vote threshold. As an example, when the vote threshold is 3, the votes which equal 3, 4, 5, 6, 7 are taken into account. But the evaluation results show that the results of 6 votes is not as good as votes equals to 5, which means that as the threshold conditions become more stringent, the correctly predicted pixels are removed. 
This phenomenon indicates that there should be a certain amount of fault tolerance in this method, because not every level of prediction is exactly right. This also proves the value of the method used in this study, our method doesn't relay on single predicted result, it integrates multi-scale information and uses multiple validation to get the optimal final result.

\section*{Conclusion}
In this paper, we have proposed a multiscale image processing method to solve the segmentation of liver cancer in histopathological images. Eight networks are compared to choose the most suitable network for liver cancer segmentation. Through the comprehensive comparison of performance, U-Net is chosen. A method of partial color normalization of hispathological images is adapted to solve the influence of background, and then seven-levels gaussian pyramid representation is built for each WSI to obtain multiscale image set. The trained U-Net is fine-tuned in each level to obtain independent model. Then, shifted cropping and weighted overlapping are adapted during prediction to solve the continuity problem of blocks. At the end, the predicted images are mapped back to the original size and a voting mechanism is proposed to combine the multiscale predicted images. We apply this method to the validation images of the 2019 MICCAI PAIP Challenge, the evaluation shows that our algorithm outperforms other algorithms from the state-of-the-art. 

This study highlighted the effectiveness of liver tumor segmentation at multiple scales. During the experiment, it was found that the prediction effect of layers is different. The reasons need to be further explored and we still investigate the characteristics of each scale. Meanwhile, it is expected to construct a new network that could integrate images at multiple scales simultaneously.

\section*{Acknowledgment}
The authors gratelfully acknowledge financial support from China Scholarship Council. Deidentified pathology images and annotations used in this research were prepared and provided by the Seoul National University Hospital by a grant of the Korea Health Technology R\&D Project through the Korea Health Industry Development Institute (KHIDI), funded by the Ministry of Health \& Welfare, Republic of Korea (grant number: HI18C0316). 

\bibliography{reference}

\end{document}